# Quantum-Enhanced Tunable Second-Order Optical Nonlinearity in Bilayer Graphene

Sanfeng Wu[1], Li Mao[2], Aaron M. Jones[1], Wang Yao[3], Chuanwei Zhang[2], Xiaodong Xu[1,4,*]

[1] Department of Physics, University of Washington, Seattle, Washington 98195, USA

[2] Department of Physics and Astronomy, Washington State University, Pullman, Washington, 99164 USA

[3] Department of Physics and Center of Theoretical and Computational Physics, The University of Hong Kong, Hong Kong, China

[4] Department of Material Science and Engineering, University of Washington, Seattle, Washington 98195, USA

**Abstract**

**Second order optical nonlinear processes involve the coherent mixing of two electromagnetic waves to generate a new optical frequency, which plays a central role in a variety of applications, such as ultrafast laser systems, rectifiers, modulators, and optical imaging. However, progress is limited in the mid-infrared (MIR) region due to the lack of suitable nonlinear materials. It is desirable to develop a robust system with a strong, electrically tunable second order optical nonlinearity. Here we demonstrate theoretically that AB-stacked bilayer graphene (BLG) can exhibit a giant and tunable second order nonlinear susceptibility $\chi^{(2)}$ once an in-plane electric field is applied. $\chi^{(2)}$ can be electrically tuned from 0 to $\sim 10^5 \text{pm/V}$, three orders of magnitude larger than the widely used nonlinear crystal $AgGaSe_2$. We show that the unusually large $\chi^{(2)}$ arises from two different quantum enhanced two-photon processes thanks to the unique electronic spectrum of BLG. The tunable electronic bandgap of BLG adds additional tunability on the resonance of $\chi^{(2)}$, which corresponds to a tunable wavelength ranging from ~2.6 μm to ~3.1 μm for the up-converted photon. Combined with the high electron mobility and optical transparency of the atomically thin BLG, our scheme suggests a new regime of nonlinear photonics based on BLG.**

**Text**

Graphene-based photonics and optoelectronics[1,2] have drawn intense interest because of their combination of unique electronic properties[3], such as high electron mobility[4] and long mean free path[5], as well as their excellent optical properties[1,2,6,7], such as broadband optical absorption[2,6,7] and ultrafast optical response[8,9,10]. In this context, BLG may have better potential than single layer graphene for photonic applications due to its four-band electronic structure and widely tunable bandgap in the MIR[11]. One promising application is to achieve chip-scale and electrically tunable nonlinear optical devices. Conventional nonlinear optical devices are based on bulk crystals which have poor compatibility with integrated circuits and lack electrical tunability in intensity and wavelength. The required phase matching conditions also lead to volatile signals subject to perturbations from the environment. All these limitations prevent the integration of nonlinear photonics and advanced electronics for new photonic and optoelectronic applications. Here we show that, under realistic device conditions, the difficulties mentioned above can be overcome using the nonlinear optical properties of BLG. Therefore BLG emerges as an excellent $\chi^{(2)}$ material and may provide opportunities for new optoelectronic and photonic applications.

Figures 1**a** and 1**b** show the atomic structure of BLG and a typical dual-gate field effect transistor (FET) device[12], respectively. We calculate the intensity of SHG following a quantum description of nonlinear optical conductivity. The applied potential bias between top and bottom gates tunes the electronic properties of BLG. The Hamiltonian of BLG near the Dirac points (***K*** and ***K'***) can be written as[3]

$$H = \sum \psi_k^\dagger \mathcal{H}_k \psi_k, \quad (1)$$

where

$$\psi_k^\dagger = (b_k^{1\dagger}, a_k^{1\dagger}, a_k^{2\dagger}, b_k^{2\dagger}),$$

and

$$\mathcal{H}_k = \begin{pmatrix} -\Delta & \hbar v_f g & 0 & 0 \\ \hbar v_f g^* & -\Delta & \gamma_1 & 0 \\ 0 & \gamma_1 & \Delta & \hbar v_f g \\ 0 & 0 & \hbar v_f g^* & \Delta \end{pmatrix}.$$

$a_k^{i\dagger}(b_k^{i\dagger})$ is the creation operator for electrons at the sublattice A (B), in the layer $i (i = 1,2)$, and with momentum ***k*** in the BZ. ***k*** $= (k_x, k_y)$ is the continuous wave vector from ***K, K'*** and $g = k_x - \xi i k_y$ is a complex number ($\xi = +1$ for ***K*** and $\xi = -1$ for ***K'***). The spatial coordinates are defined in Fig. 1**a** and 1**b**. $v_f \approx 10^6$ m/s is the Fermi velocity of single layer graphene. $\Delta (-\Delta)$ describes the potential effect induced by the gate voltages. $\gamma_1 \approx 0.4$ eV is the inter-layer hopping parameter[3]. Here we ignore other hopping processes due to their relatively weak strengths. The Hamiltonian (1) has four energy bands, which are obtained by diagonalizing $\mathcal{H}_k$, and plotted in Figure 1**c** for $\Delta = 0$.

The total Hamiltonian $\mathcal{H}_p$ that includes the interactions between the BLG and normal incident MIR photons can be obtained by replacing the momentum vector $\hbar k$ in $\mathcal{H}_k$ with $p = \hbar k + eA$, where ***A*** is the vector potential of the optical fields[3,13,14,15]. The evolution of the electronic state is determined by the quantum Liouville equation[16]:

$$i\hbar \partial_t \rho = [\mathcal{H}_p, \rho] - i\Gamma(\rho - \rho_{t=0}), \quad (2)$$

where $\rho$ is the time dependent quantum state of electrons with momentum ***k*** at temperature $T$ and chemical potential $\mu$. $\Gamma$ is a phenomenological parameter for describing the relaxation of electronic states. We set $\Gamma = 0.05$ eV in our calculations[17], which corresponds to ~80 fs relaxation time of electrons. This is a conservative estimation even at room temperature (The effect of $\Gamma$ is discussed in supplemental materials[18]). The steady state density matrix in the presence of the optical fields can be obtained by solving the time-dependent equation (2) perturbatively[16] (see supplemental materials[18]).

The optically induced electric current is defined as[19,20,21]

$$j_s = ve \sum_{BZ} tr(\rho \frac{\partial \mathcal{H}_p}{\partial p}) = \sigma_1 E e^{i\omega t} + \sigma_2 E^2 e^{i2\omega t} + c.c + \cdots, \quad (3)$$

where $v = 2$ describes the spin degeneracy and $E$ is the electric field. $\sigma_1$ is the linear optical conductivity, which value obtained from our approach is exactly the same as the ones reported using the Kubo formula[7]. Here we focus on the second order nonlinear optical conductivity $\sigma_2$, which is responsible for SHG. The simplified formula of $\sigma_2^\alpha$ is[18]:

$$\sigma_2^\alpha = \frac{e}{2\pi^2 E^2} \int_{BZ} \sum_{ijl} \left[ \frac{\rho_k^{ii} - \rho_k^{ll}}{\hbar\omega + \epsilon_i - \epsilon_l - i\Gamma} - \frac{\rho_k^{ll} - \rho_k^{jj}}{\hbar\omega + \epsilon_l - \epsilon_j - i\Gamma} \right] \frac{(H_{int})_{il} (H_{int})_{lj} (\eta^\alpha)_{ji}}{2\hbar\omega + \epsilon_i - \epsilon_j - i\Gamma} dk, \quad (4)$$

where $\alpha = x, y$; $\eta^\alpha = \frac{\partial \mathcal{H}_p}{\partial p_\alpha}$; $H_{int} = \mathcal{H}_p - \mathcal{H}_k$ is the interaction Hamiltonian and $\rho_k = \text{diag}[f_1, f_2, f_3, f_4]$ is the initial state obtained from the Fermi-Dirac distribution $f_i = \frac{1}{\exp[(\epsilon_i - \mu)/k_B T] + 1}$ of electrons at different bands with the band energy $\epsilon_i (i = 1,2,3,4)$. $\mu$ is the chemical potential, $k_B$ is the Boltzmann constant, and $T$ is the temperature.

The above equation (4) has a trivial solution $\sigma_2^\alpha = 0$, although the inversion symmetry of intrinsic BLG ($C_{3v}$ group) has been broken by the vertical biased potential. This is due to the cancellation of the optically-excited second order electrical current at opposite momenta ($+k$ and $-k$) with respect to both Dirac points. In order to generate a non-zero second order optical nonlinearity, we introduce an in-plane electrical current between source and drain to suppress the cancellation. From semi-classical electron transport theory, the in-plane electric field shifts the Fermi surface by $\Delta k = \frac{m^* u_m j_{sc}}{\hbar \sigma_{dc}}$, where $m^*$ is the electronic effective mass, $j_{sc}$ is the in-plane current density, $\sigma_{dc}$ is the DC conductivity, and $u_m$ is the mobility[18] (Fig. 1c). According to experimental data of BLG[1,2,22,23], we take $m^* \approx 0.05 m_e$, $u_m \approx 1 \frac{m^2}{Vs}$, $\sigma_{dc} \approx \frac{e^2}{\hbar}$. An energy shift of $\hbar v_f |\Delta k| \approx 0.01$ eV is achievable with a DC current $j_{sc} \approx 8$ nA/nm. As a result, the Fermi level is shifted (Fig. 1c) and the initial electronic state in the presence of $j_{sc}$ turns out to be $\rho_k \leftarrow \rho_{k-\Delta k}$. Physically, due to the Pauli exclusion principle, part of the optical transitions are blocked due to the Fermi level shifting. Therefore the $+k/-k$ symmetry is broken and SHG becomes nonzero.

We first explore the second order nonlinear optical conductivity with circularly polarized light excitations. For simplicity, we use left circular polarization as an example. Figures 2a and 2b plot both the real and imaginary parts of the second order optical conductivity $\sigma_2$ at temperatures 30 K and 300 K, respectively. $\sigma_2$ is decomposed into $x$ and $y$ directions. The induced current at twice the optical frequency has a strong phase-correlation between the $x$ and $y$ directions: $\sigma_2^y = -i\sigma_2^x$. It therefore implies the left circular polarization of the new optical fields generated by SHG are the same as the incident beam. We set the chemical potential $\mu = \gamma_1/2 = 0.2$ eV without the bandgap opening (i.e., $\Delta = 0$). The in-plane electric field is along the y direction, leading to a shift of $\hbar v_f |\Delta k| \sim 0.01$ eV. A surprising finding here is the giant enhancement of $\sigma_2$ at $\omega = 0.2$ eV and $0.4$ eV, as seen from the resonant peaks in Figures 2a and 2b. The intensities of the resonant peaks increase as the temperature decreases. The ratio of $\sigma_2$ at low temperature (30K) to high temperature (300K) is about 1.5. In Fig. 2c, we plot $\sigma_2$ as a function of both temperature and frequency, where these features can be clearly seen.

The two resonant peaks arise from different quantum-enhancement mechanisms. At $\omega = 0.4$ eV the optical process is dominated by the transitions from the 2$^{nd}$ to the 4$^{th}$ band (Fig. 2d) through two-photon absorption. Although two-photon absorption can happen even in BLG in equilibrium[24,25], the contributions to SHG from the $K + k$ and $K - k$ points cancel out. As we mentioned above, the in-plane electric field breaks this symmetry, leading to a shift of the Fermi surface to avoid such cancellation. More importantly, when $\omega = \Delta E_{23} = \Delta E_{34}$ ($\Delta E_{ij}$ is the energy difference between the bands $i$ and $j$), the transient state of SHG is a real state (i.e. the electronic state in the 3$^{rd}$ band), giving rise to resonantly enhanced contribution as compared to conventional SHG where the intermediate states are virtual states. In this case, the so-called double resonance enhanced (DRE) harmonic generation happens in the BLG system. The signal intensity of DRE SHG[26] is usually proportional to $1/\Gamma^2$.

At $\omega = 0.2$ eV, the high joint density-of-states (HJDOS) come into play. As we see from Figure 2d, the energy splitting $\Delta E_{34} \approx \gamma_1$ is nearly a constant for a large range of $k$, which allows plenty of resonant 2-photon transitions when the incident laser frequency $\omega \sim \frac{\gamma_1}{2} = 0.2$ eV, giving rise to a significant enhancement of SHG. Clearly, the enhancement of SHG at both $\omega = 0.4$ eV and $0.2$ eV originates from the special band structure of the BLG.

We now explore the second order nonlinear optical conductivity with linearly polarized light excitations. Figure 3 shows $\sigma_2$ at the two quantum-enhanced resonant peaks ($\omega = 0.2$eV and $0.4$eV) as a function of the incident polarization angle $\varphi$ with respect to the $x$-axis. The temperature is set at 180K. The red solid line denotes the nonlinear response along the x direction, with the blue dashed line along the y direction. We note that the induced current in the two directions have almost the same phase, which indicates that the output SHG is still linearly polarized. When we vary the incident angle, the optical conductivities at two different resonant frequencies behave differently. (1) $\sigma_2$ enhanced by HJDOS has a polarization angle $2\varphi + \frac{\pi}{2}$ (Fig. 3a). For instance, when the incident laser is polarized at $\varphi = \frac{\pi}{4}$, the second order nonlinear optical conductivity is polarized with angle $\pi$, i.e. in the $x$ direction. (2) The optical current excited by the DRE mechanism has a preferred axis ($y$-axis) (Fig. 3b). The signal generated in the $y$ direction (where the in-plane electric field lies) is larger than that in the $x$ direction. Therefore the output SHG intensity in this situation oscillates around the $y$-axis as $\varphi$ varies.

Here we note that the symmetric patterns in Fig. 3 do not depend on the relative direction of the in-plane electric fields and lattice orientation. For simplicity, we here set the applied bias along the lattice direction (-y), which is not necessary in practice. In general, we find that the second order optical response always behaves in the same way as that shown in Fig. 3 for varying directions of the DC current, which determines the symmetric axis of the polar plots (supplementary materials).

The second order optical susceptibility $\chi^{(2)}$ can be estimated from the optical conductivity[19]:

$$\chi^{(2)} = \frac{\sigma_2}{\omega \, \varepsilon_0 d_{BLG}}. \quad (5)$$

Here $d_{BLG} \sim 0.66$ nm is the thickness of BLG and $\varepsilon_0 = 8.85 \times 10^{-12}$ C/Vm is the vacuum permittivity. Using $ev_f/V^2$ as a unit for $\sigma_2$, we have $\chi^{(2)} \sim 10^5$ pm/V, about three orders of magnitude larger than state-of-the-art nonlinear crystals in MIR. For comparison, the second order nonlinear susceptibility of the widely used AgGaSe$_2$ crystal is $68 \text{ pm/V}$ at a wavelength of ~$2.1 \text{ um}$[27]. At low temperature and with a large in-plane electric field, the value can be even larger and the response region in our scheme is ultra-broad, as shown below.

We now show that the giant SHG can be tuned by the in-plane electric field, Fermi level, and biased potential $\Delta$. Fig. **4a** plots the nonlinear optical conductivity at two resonant peaks as a function of the in-plane current, which is estimated from the shift of the Fermi surface ($\Delta \boldsymbol{k}$). As an example, the incident laser is set to be polarized in the *x* direction and $j_{sc}$ is in the *y* direction at a temperature of 180 K. The Fermi level is set to be $0.2 \text{ eV}$. The red stars in the figure represent the parameters we used in the previous discussions in this paper. This plot indicates that SHG intensity increases linearly when the in-plane current is small (< ~35 nA/nm), and saturates at high in-plane current. Apparently, if we turn off the in-plane field, SHG is consequently switched off. Unlike conventional electric field-induced SHG which requires an extremely large electric voltage[28] (kilovolts) in the crystal, small in-plane electric fields are more favorable in device applications. In BLG, with an electric field of ~1 V/mm, one can electrically switch on and off SHG.

The SHG signal can also be tuned by adjusting the Fermi level, as shown in Fig. **4b**. Obviously, there is a pronounced SHG when the Fermi level falls in the range around $\frac{\gamma_1}{2} = 0.2 \text{eV}$ with $\Delta = 0$. If we increase or decrease the Fermi level without crossing the Dirac point, the effects of the DRE and HJDOS become less important and the signal decreases to zero. One may notice that the SHG signal will recover when the Fermi level goes down to $-0.2$ eV (in the p-doping situation).

Finally, we show that the unique bandgap tunability of BLG by an interlayer bias induces the widely tunable resonant frequency of SHG. The intensity plot in Fig. **4c** shows the magnitude of the second order optical conductivity as a function of $\Delta$ and incident laser frequency. In addition to the change of the amplitude of $\sigma_2$, an important effect is the variation of the resonant frequency $\omega_0$, indicated in Fig. **4c** by the dashed lines. $\omega_0$ increases from 0.4 to ~0.48 eV for the DRE peak and from 0.2 eV to ~0.25 eV for the HJDOS peak, as $\Delta$ is tuned from 0 to 0.15 eV. Therefore the resonant wavelengths for SHG are electrically tunable from ~5 μm to ~6 μm and from ~2.6 μm to ~3.1 μm. This is another remarkable advantage as compared to conventional nonlinear crystals, where the wavelength for SHG can only be changed by rotating the crystal angles. Moreover, due to its atomic thickness, BLG requires no phase matching condition, which is crucial in conventional SHG. Phase matching usually leads to increased sensitivity to environmental perturbations, such as temperature. Therefore, from this point of view, SHG of BLG is robust.

In conclusion, we propose a new scheme based on BLG as a nonlinear optical material, with an extremely large second order optical susceptibility $\chi^{(2)} \sim 10^5 \text{ pm/V}$. We find that this enhancement arises from two different types of quantum-enhanced mechanisms, unique to the electronic structure of BLG. Our calculation shows an excellent electrical tunability of the optical nonlinearity in both intensity and wavelength compared to conventional nonlinear crystals. Considering graphene's many unique properties, we believe that our results encourage further experimental investigation towards nonlinear photonics based on BLG and may open new opportunities for graphene applications.

**Corresponding Author**

[*]Email: xuxd@uw.edu

**Acknowledgements:**

This work is supported by DARPA YFA N66001-11-1-4124. L. Mao and C. Zhang are supported by DARPA-YFA N66001-10-1-4025, DARPA-MTO (FA9550-10-1-0497), and NSF-PHY (1104546). A. Jones is supported by NSF Grant No. DGE-0718124. W. Yao is supported by Research Grant Council of Hong Kong.

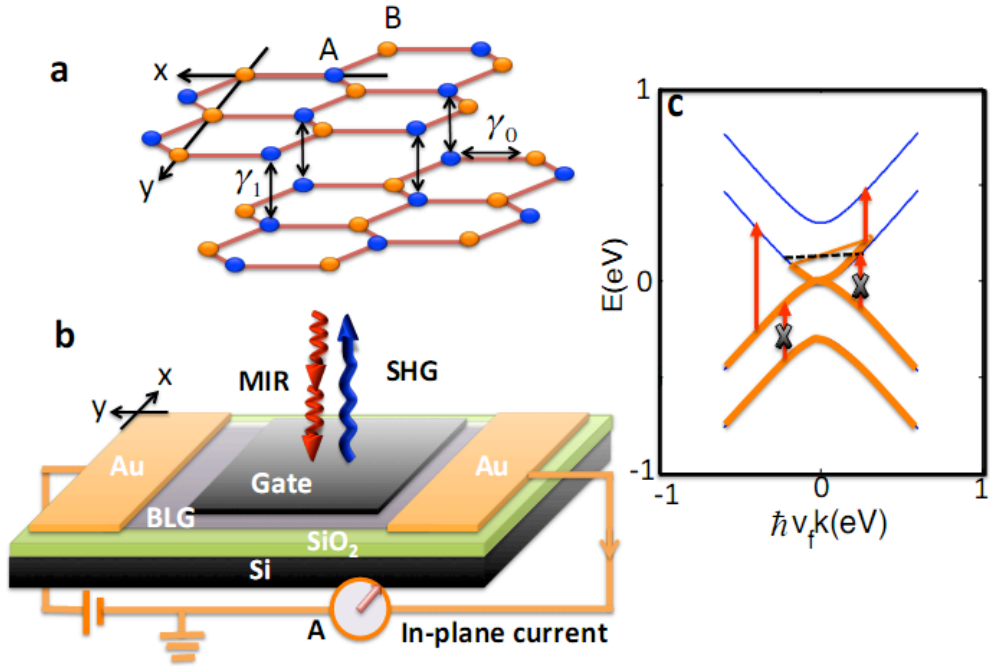

**Figure 1.** Dual-gated bilayer graphene FET and its optical transitions. (a) Atomic structure of AB-stacked bilayer graphene. $\gamma_1$ and $\gamma_0$ are the inter- and intra- layer hopping parameters. The coordinates used in the calculations are shown. (b) Experimental schematic of using a dual-gated bilayer graphene FET to realize SHG. The in-plane current is set in the –y direction in our calculation. (c) Energy spectrum of BLG and Fermi level shifting when there is an in-plane current between source and drain. Possible and forbidden optical transitions are shown as red arrows. Brown color on the energy levels denotes the occupied states.

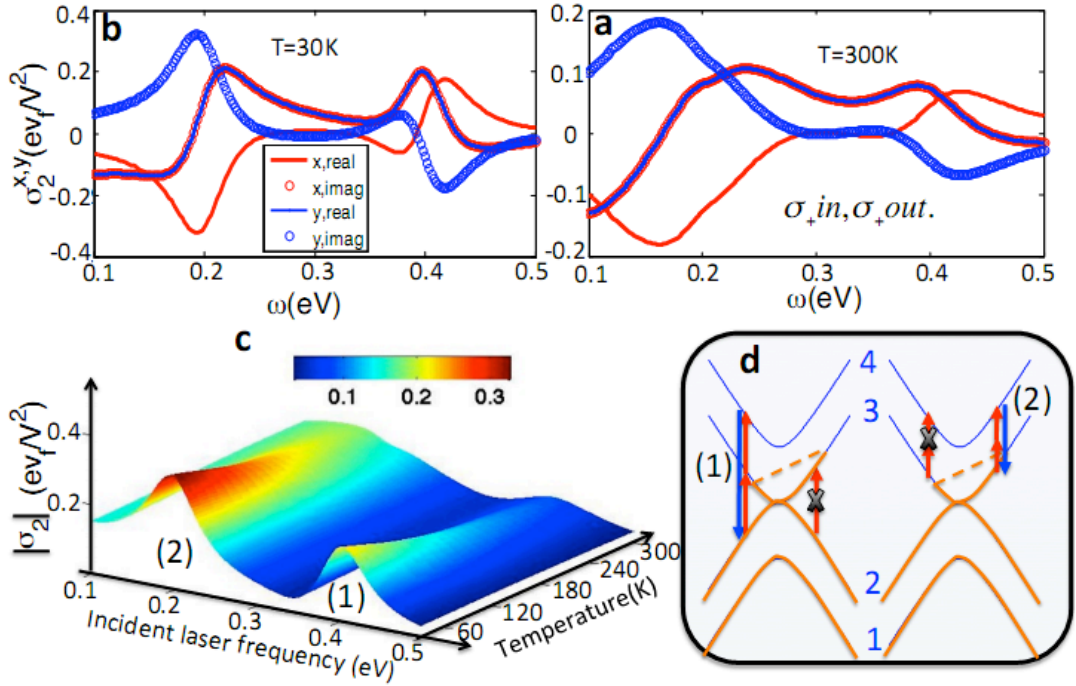

**Figure 2.** Second order optical response of BLG under left-hand circularly polarized incident laser beam. (a) and (b), frequency dependent nonlinear optical conductivity $\sigma_2$ calculated at a temperature of 30K and 300K, respectively. The interlayer hopping parameter $\gamma_1 = 0.4$ eV and the phenomenal energy broadening $\Gamma$ is set to $0.05\ eV$. Both real and imaginary parts are shown on both x and y directions. The plots show that there are strongly enhanced responses located at $\omega = 0.2$ eV and $0.4$ eV. (c) $\sigma_2$ as a function of temperature and laser frequency. (d) Corresponding optical process for the response peak at (1) $\omega = 0.4$ eV and (2) 0.2 eV.

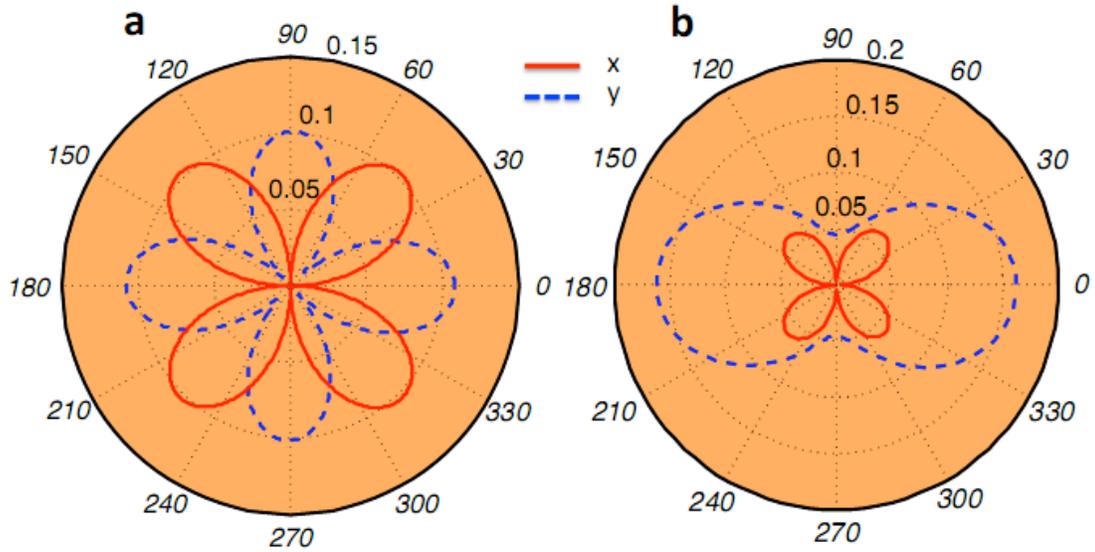

**Figure 3.** Second order optical response of BLG under linearly polarized incident laser beam. Polar plots of $\sigma_2$ versus varying polarization angles of the incident beam starting from the *x*-axis at T=180 K for (a), $\omega = 0.2$ eV and (b), $\omega = 0.4$ eV. The red solid and blue dashed curves correspond to the response in the x and y directions, respectively.

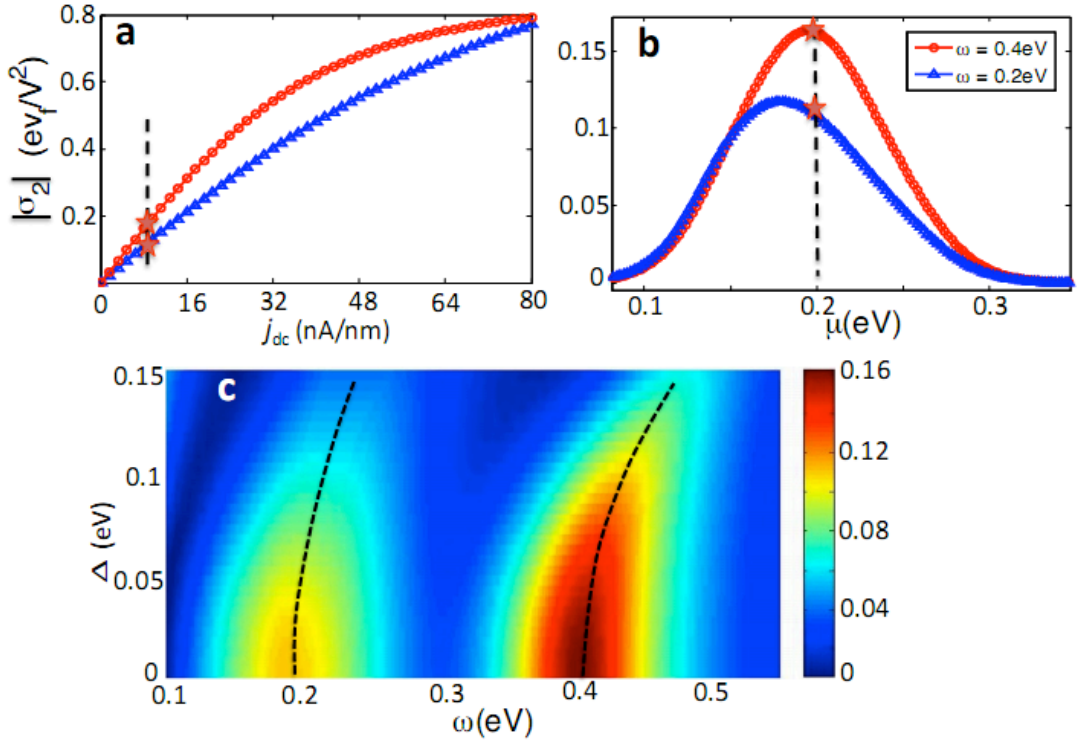

**Figure 4.** Tunable effects of SHG of BLG. (a) Amplitude of $\sigma_2$ as a function of estimated in-plane current ($\mu = 0.2\,eV$). (b) Amplitude of $\sigma_2$ as a function of Fermi level $\mu$. The red star denotes the value used in our previous calculations. (c) Tunable SHG by controlling the electronic bandgap of BLG. The black dashed lines denote the resonant frequencies. Temperature is set to 180 K.

**Supplemental Materials**

**Quantum-Enhanced Tunable Second-Order Optical Nonlinearity in Bilayer Graphene**

Sanfeng Wu[1], Li Mao[2], Aaron M. Jones[1], Wang Yao[3], Chuanwei Zhang[2], Xiaodong Xu[1,4*]

[1] Department of Physics, University of Washington, Seattle, Washington 98195, USA
[2] Department of Physics and Astronomy, Washington State University, Pullman, Washington 99164 USA
[3] Department of Physics and Center of Theoretical and Computational Physics, The University of Hong Kong, Hong Kong, China
[4] Department of Material Science and Engineering, University of Washington, Seattle, Washington 98195, USA
*Email: xuxd@uw.edu

**1. Theory for second order optical conductivity**

Considering the applied potential bias between top and bottom gates, the electronic Hamiltonian near the Dirac points (***K*** and ***K'***) is[1]:

$$H = \sum \psi_k^\dagger \mathcal{H}_k \psi_k,$$

where

$$\psi_k^\dagger = (b_k^{1\dagger}, a_k^{1\dagger}, a_k^{2\dagger}, b_k^{2\dagger}),$$

and

$$\mathcal{H}_k = \begin{pmatrix} -\Delta & \hbar v_f g & 0 & 0 \\ \hbar v_f g^* & -\Delta & \gamma_1 & 0 \\ 0 & \gamma_1 & \Delta & \hbar v_f g \\ 0 & 0 & \hbar v_f g^* & \Delta \end{pmatrix},$$

$a_k^{i\dagger}(b_k^{i\dagger})$ is the creation operator for electrons at the sublattice A (B), in the layer $i(i = 1,2)$, and with momentum k in the BZ. $k = (k_x, k_y)$ is the continuous wave vector from K, K' and $g = k_x - \xi i k_y$ is a complex number ($\xi = +1$ for K and $\xi = -1$ for K'). The Fermi velocity $v_f$ is determined by the intra-layer hopping energy $\gamma_0 \approx 2.8\ eV$ between nearest neighbors. $a$ is the lattice constant of graphene. $\gamma_1 \approx 0.4\ eV$ is the inter-layer hopping parameter between $A_1$ and $A_2$. We ignore other hopping processes due to their relatively weak strengths. The diagonal items $\pm\Delta$ come from the potential bias between the two graphene-layers. Diagonalizing the Hamiltonian, we can obtain the energy spectrum with four branches near the Dirac point[1].

The interaction between light and BLG can be described by replacing the momentum vector $\hbar \mathbf{k}$ with[1]

$$\mathbf{p} = \hbar \mathbf{k} + e\mathbf{A},$$

where $\mathbf{A} = -\mathbf{E}/(i\omega)$ is the vector potential of the incident laser beam. As an example, if the incident laser is a circularly polarized beam $\sigma_+$, then $\mathrm{E} = \frac{E e^{i\omega t}}{2i}(1,-i) + \mathrm{c.c.}$ The interaction between laser and electrons (~1meV) is weak compared to the system energy scale (~0.1eV), therefore we can write the total Hamiltonian under band representation as
$$\bar{\mathcal{H}}_p = \bar{\mathcal{H}}_k + \frac{ev_f E e^{i\omega t}}{2\omega} H_\Phi.$$
where $H_\Phi = \Phi^\dagger[I\otimes(\sigma_x - i\xi\sigma_y)]\Phi$ for a circularly polarized beam $\sigma_+$. $I$ is a $2 \times 2$ unit matrix. $\sigma = (\sigma_x, \sigma_y)$ is the vector of Pauli matrixes. $\Phi$ is the initial eigenstate without the laser-field satisfying $\bar{\mathcal{H}}_k = \Phi^\dagger \mathcal{H}_k \Phi = \text{diag}[\epsilon_1, \epsilon_2, \epsilon_3, \epsilon_4]$, where $\epsilon_i (i = 1,2,3,4)$ is the energy of the i$^{th}$ band.

The evolution of electronic states is determined by the quantum Liouville equation[2,3]
$$i\hbar\partial_t \rho = [\bar{\mathcal{H}}_p, \rho] - i\Gamma(\rho - \rho(t=0)),$$
where $\rho$ is the time dependent quantum state of electrons with momentum $\mathbf{k}$ at temperature $T$ and chemical potential $\mu$. $\Gamma$ describes the electronic relaxation time[4,5]. The linear and nonlinear optical response of $\rho$ to the optical fields can be obtained by solving the Liouville equation perturbatively
$$\rho = \rho_k + [\rho_1(t)e^{i\omega t} + c.c]E + [\rho_2(t)e^{i2\omega t} + c.c]E^2 + \cdots.$$
The initial state $\rho_k = \text{diag}[f_1, f_2, f_3, f_4]$ is the initial state obtained from the Fermi-Dirac distribution $f_i = \frac{1}{\exp[(\epsilon_i - \mu)/k_B T]+1}$ of electrons.

As we mentioned in the main text, the SHG does not exist unless there is an in-plane electric field. According to semi-classical electron transport theory, if there is an in-plane electric field $\mathcal{E}$, the Fermi surface is shifted by wavenumber $\Delta \mathbf{k} = e\tau\mathcal{E}/\hbar = m^* v_d/\hbar$ along the direction opposite the electric field owing to the negative electron charge. $\tau$ is the relaxation time, $m^*$ is the electronic effective mass and $v_d$ is the drift velocity. This process is responsible for the leak current $j_{sc} = \sigma_{dc}\mathcal{E}d$ measured by the ammeter, where d is the sample length perpendicular to the electric field, assuming a rectangular sample shape. Replacing $v_d$ with the electronic mobility $u_m = |v_d/\mathcal{E}|$, we have
$$\hbar\Delta \mathbf{k} = \frac{m^* u_m j_{sc}}{\sigma_{dc} d}.$$
According to experimental data of BLG[6,7], we choose $m^* \approx 0.05 m_e, u_m \approx 1\frac{m^2}{Vs}, \sigma_{dc} \approx \frac{e^2}{\hbar}, v_f \approx 10^6 m/$, then $\hbar v_f \Delta \mathbf{k} \approx 0.01 eV$ for a leak current $j_{sc} \approx 8\ \mu A/\mu m$ (corresponding to $\mathcal{E} \approx 2000 V/m$).

As a result of the shift of the Fermi surface, the initial electronic state in the presence of the in-plane applied current turns out to be
$$\rho_k \leftarrow \rho_{k-\Delta k}.$$
Substituting this into the quantum Liouville equation, we obtain the first order and second order equations for the electronic-state evolution
$$i\hbar\partial_t \rho_1 - (\hbar\omega - i\Gamma)\rho_1 = [\bar{\mathcal{H}}_k, \rho_1] + \frac{ev_f}{2\omega}[H_\Phi, \rho_k],$$
$$i\hbar\partial_t \rho_2 - (2\hbar\omega - i\Gamma)\rho_2 = [\bar{\mathcal{H}}_k, \rho_2] + \frac{ev_f}{2\omega}[H_\Phi, \rho_1].$$
The dynamics of the electronic state excited by photons can be obtained by solving these two linear ordinary differential equations as a steady state problem (*i.e.*, $\partial_t \rho_1 = 0, \partial_t \rho_2 = 0$), yielding

$$\rho_1{}^{ij} = -\left(\frac{ev_f}{2\omega}\right)\frac{[H_\Phi, \rho_k]_{ij}}{\hbar\omega + \epsilon_{ij} - i\Gamma},$$

$$\rho_2{}^{ij} = -\left(\frac{ev_f}{2\omega}\right)\frac{[H_\Phi, \rho_1]_{ij}}{2\hbar\omega + \epsilon_{ij} - i\Gamma}.$$

The induced electric current is defined as

$$j_e = v e \sum_{BZ} \text{tr}(\rho \frac{\partial \bar{\mathcal{H}}_p}{\partial \boldsymbol{p}}) = \sigma_1 E e^{i\omega t} + \sigma_2 E^2 e^{i2\omega t} + c.c + \cdots.$$

where $v = 2$ describes the spin degeneracy and $\sigma_1$ is the linear optical conductivity. We have confirmed that our result for $\sigma_1$ is exactly the same as that in former literature on BLG using the Kubo formula. Here we focus on the second order nonlinear optical conductivity $\sigma_2$, which stimulates SHG. The result is

$$\sigma_2^\alpha = 2e \sum_{BZ} tr\left(\rho_2 \frac{\partial \bar{\mathcal{H}}_p}{\partial p_\alpha}\right) = 2ev_f \sum_{K,K'} tr(\rho_2 \eta_\Phi^\alpha)$$

$$= \frac{e^3 v_f^2}{8\pi^2 \omega^2} \int_{BZ} \sum_{ijl} \left[\frac{\rho_k{}^{ii} - \rho_k{}^{ll}}{\hbar\omega + \epsilon_i - \epsilon_l - i\Gamma} - \frac{\rho_k{}^{ll} - \rho_k{}^{jj}}{\hbar\omega + \epsilon_l - \epsilon_j - i\Gamma}\right] \frac{(H_\Phi)_{il}(H_\Phi)_{lj}(\eta_\Phi^\alpha)_{ji}}{2\hbar\omega + \epsilon_i - \epsilon_j - i\Gamma} d\boldsymbol{k}.$$

Here $\eta_\Phi^\alpha = \frac{\partial \mathcal{H}_p}{\partial p_\alpha} = v_f \Phi^\dagger (I \otimes \sigma_\alpha) \Phi$ for Dirac cone K and $\eta_\Phi^\alpha = v_f \Phi^\dagger (I \otimes \sigma_\alpha^*) \Phi$ for K'. $i, j, l = 1, 2, 3, 4$ and the integration over $k$ is performed in the vicinity of K or K'. $\alpha = x, y$ represents the axis direction in the graphene plane.

**2. Effects of Γ**
We set $\Gamma = 0.05\text{eV}$ in our calculation in the main text. The value of Γ affects the magnitude of the second order optical conductivity significantly. In Fig. 1S. we show $\sigma_2$ as a function of Γ and the incident laser frequency ω. The incident laser is polarized along the x direction and T=30K. Apparently, the signal intensity becomes larger when the relaxation time of the excited electronic state becomes longer, i.e., smaller Γ.

**3. Effects of the direction of the in-plane DC current**
To achieve SHG, we have to apply an in-plane electric field. In the main text, we apply an in-plane current along the y direction as an example. In general, the direction of the field is not unique. One can obtain giant optical nonlinear conductivity by inducing a current in any direction in the 2D atomic plane. This is important in practice because the in-plane current may not lie in the y direction defined by the atomic structure (See Fig.1 in the main text). Here we show the results for a different direction ($-y'$) of the in-plane current with a linearly polarized incident laser at T=30 K (See Fig. 2S). We show that it is the direction of the DC current that determines the symmetric axis for second order optical conductivity (Fig. 2S **c** and **d**).

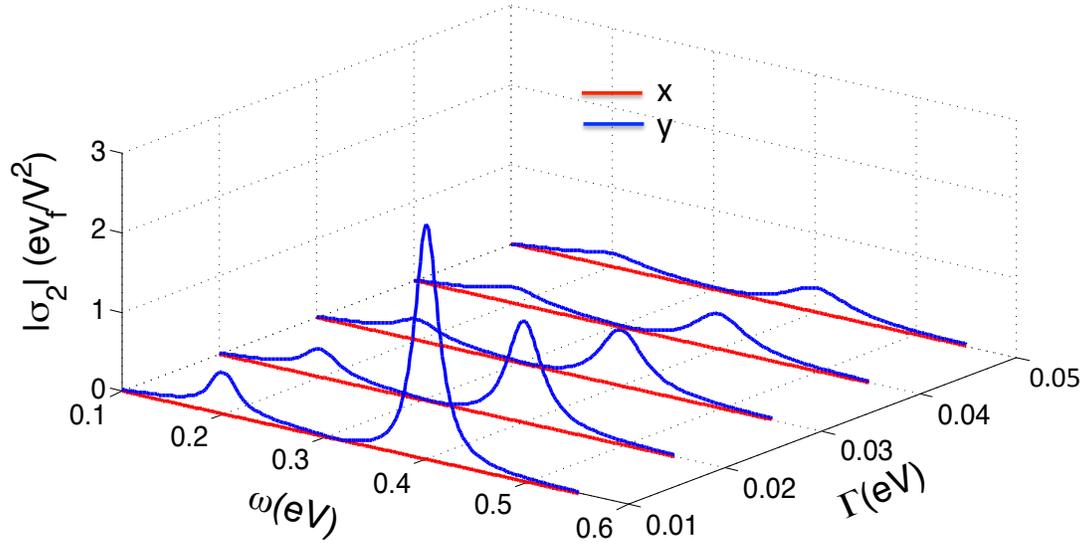

Figure 1S | Effects of relaxation time of excited states on SHG. The incident laser is linearly polarized and T=30 K. Blue line denotes the second order optical conductivity in the y direction and red in the x direction.

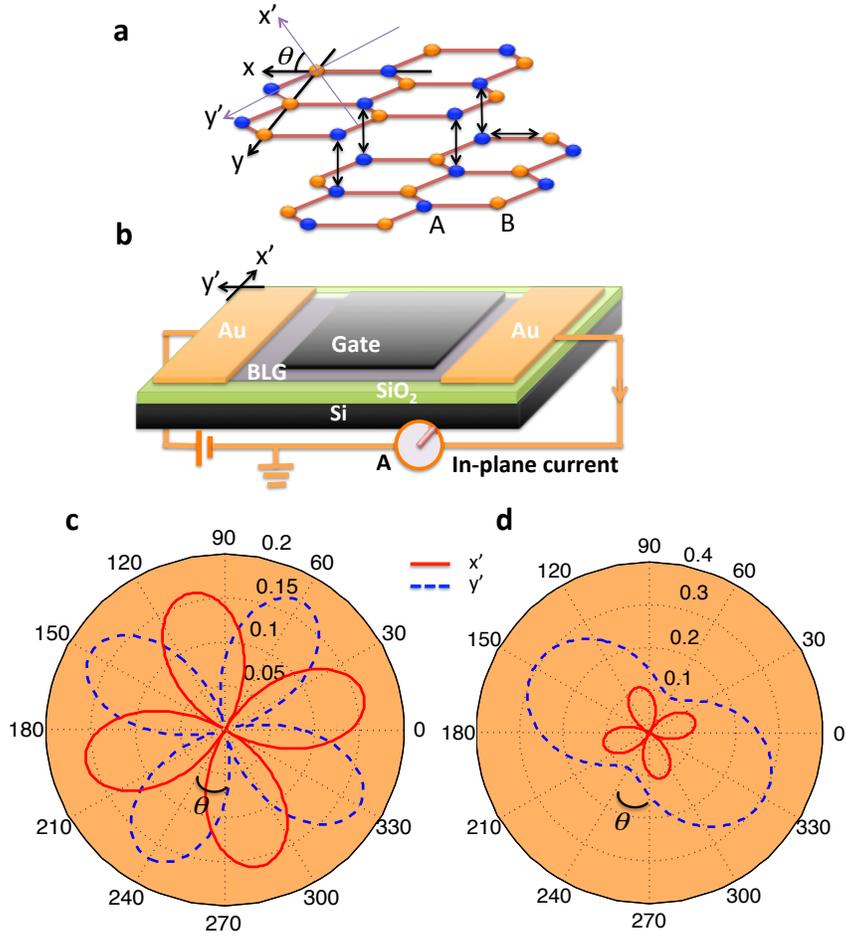

Figure 2S | Effect of the direction of in-plane electric field on SHG. **a**, Atomic structure of BLG. (x,y) and the coordinates used for tight-binding model. ($-y'$) is the direction of the DC current as shown in **b**. **b**, Normal device scheme of BLG, which determines the direction of the current, which is usually not along the axis of the coordinates (x,y) in **a**. The deviation is described by an angle θ. **c** and **d**, polar plots of the intensity of the second order optical conductivity in the x' and y' direction at T=30 K for incident laser frequency (**c**) $\omega = 0.2$ eV and (**d**) $\omega = 0.4$ eV. We can see that the direction of the in-plane field determines the symmetric axis (shown by θ) of the optical nonlinearity. Polar angle is the polarization angle of the incident beam starting from the x-axis.